\documentclass[prd,aps,a4paper,floatfix,amsmath,amssymb,twocolumn,nofootinbib]{revtex4-1}
\usepackage{graphicx,color,array,float,fullpage}
\newcommand{\blue}{}
\DeclareMathOperator{\tr}{tr}

\begin{document}

\title{Critical phenomena in gravitational collapse with two competing
  massless matter fields} \author{Carsten Gundlach}
\affiliation{Mathematical Sciences, University of Southampton,
  Southampton SO17 1BJ, United Kingdom} \author{Thomas W. Baumgarte}
\affiliation{Department of Physics and Astronomy, Bowdoin College,
  Brunswick, ME 04011, USA} \author{David Hilditch}
\affiliation{CENTRA, Departamento de F\'isica, Instituto Superior
  T\'ecnico IST, Universidade de Lisboa UL, Avenida Rovisco Pais 1,
  1049 Lisboa, Portugal} \date{16 August 2019, revised 8 October 2019}

\begin{abstract}

In the gravitational collapse of matter beyond spherical symmetry,
gravitational waves are necessarily present. On the other hand,
gravitational waves can collapse to a black hole even without
matter. One might therefore wonder whether the interaction and
competition between the matter fields and gravitational waves affects
critical phenomena at the threshold of black hole formation. As a toy
model for this, we study type II critical collapse with two matter
fields in spherical symmetry, namely a scalar field and a Yang-Mills
field. On their own, both display discrete self-similarity (DSS) in
type II critical collapse, and we can take either one of them as a toy
model for gravitational waves. To our surprise, in numerical time
evolutions we find that, for sufficiently good fine-tuning, the scalar
field always dominates on sufficiently small scales. We explain our
results by the conjectured existence of a ``quasi-discretely
self-similar'' (QSS) solution shared by the two fields, equal to the
known Yang-Mills critical solution at infinitely large scales and the
known scalar field critical solution (the Choptuik solution) at
infinitely small scales, with a gradual transition from one field to
the other. This QSS solution itself has only one unstable mode, and so
acts as the critical solution for any mixture of scalar field and
Yang-Mills initial data.

\end{abstract}

\maketitle

%%%%%%%%%%%%%%%%%%%%%%%%%%%%%%%%%%%%%%%%%%%%%%%%%%%%%%%%%%%%%%%%%%%%%%%%%%%%%%%

\section{Introduction}

%%%%%%%%%%%%%%%%%%%%%%%%%%%%%%%%%%%%%%%%%%%%%%%%%%%%%%%%%%%%%%%%%%%%%%%%%%%%%%%

In simple self-gravitating systems in spherical symmetry, such as a
scalar field \cite{Choptuik1993} or ultrarelativistic perfect fluid
\cite{EvansColeman}, the time evolution of initial data which are
close to the threshold of black hole formation, but otherwise generic,
displays certain universal features which are called (type II) critical
phenomena in gravitational collapse. In particular, on the
supercritical (black hole-forming) side of the threshold, the black
hole mass becomes arbitrarily small as the threshold is approached,
while on the subcritical (dispersing) side of the threshold, the
maximal curvature that occurs during the evolution becomes arbitrarily
large. Let $p$ be any smooth parameter of the initial data such that, for
all other parameters held fixed, a black hole is formed precisely for
$p>p_*$. Then the black hole mass scales as $(p-p_*)^\gamma$, and the
maximum curvature as $(p_*-p)^{-2\gamma}$, for some universal critical
exponent~$\gamma>0$.

Critical phenomena in gravitational collapse in such simple systems
are explained by the existence of a critical solution, which
has the key properties that it is regular, self-similar, and has
precisely one unstable perturbation mode. This solution appears as an
intermediate attractor in the time evolution of near-critical initial
data. The effect of fine-tuning $p$ to its critical value $p_*$ in any
one-parameter family is to tune the amplitude of the one unstable mode
to zero. In the limit of perfect fine-tuning, the critical solution
persists to arbitrarily small scales and correspondingly large
curvature, and a naked singularity is formed. Hence type II critical
collapse, at least in the known examples in spherical symmetry,
indicates strongly that cosmic censorship holds only for generic
initial data, but is violated by a set of initial data that in some
sense has codimension one (see, e.g., \cite{GundlachLRR}
for a review).

It is of great interest to explore whether the critical collapse
scenario continues to hold beyond spherical symmetry, and in
particular in the presence of angular momentum or in the collapse of
gravitational waves in vacuum.

Going beyond spherical symmetry, we have investigated critical
phenomena in gravitational collapse for rotating
\cite{BaumgarteGundlach} and non-rotating \cite{CelestinoBaumgarte}
perfect fluids in axisymmetry. The critical collapse of an
axisymmetric scalar field has been investigated in
\cite{ChoptuikHirschmannLieblingPretoriusscalar} and
\cite{Baumgartescalar}, see also \cite{Laguna,Deppe,Clough}. Critical
phenomena in axisymmetric vacuum collapse were reported in
\cite{AbrahamsEvans}, and partly confirmed much more recently in
\cite{HWB}, see also \cite{Alcubierre, GarfinkleDuncan, Rinne, Sorkin,
  DHTWB13} for other attempts in the interval between these two
papers.

The vacuum case is the most interesting, since it mostly cleanly
displays the matter-independent properties of general relativity --
but it also appears to be the most difficult numerically. This brings
us back to axisymmetric matter models. In these, however,
gravitational waves are always also necessarily present. Therefore,
critical collapse cannot be moderated by a single critical
solution. The obvious conservative scenario has two separate critical
solutions dominated by the matter content and by gravitational waves,
respectively. But more exotic scenarios are possible, and one of these
will be illustrated in this paper.

As a toy model for {\em
  axisymmetric} systems with two wave-like degrees of freedom (matter
and gravitational waves), we investigate here a system with two
wave-like matter degrees of freedom in {\em spherical} symmetry,
namely a massless minimally coupled scalar field and a Yang-Mills
field, coupled to each other only gravitationally.

The critical solution and the critical exponent $\gamma$ for each
system on its own are known. Both critical solutions are {\em
  discretely} self-similar (from now on, DSS), meaning they are
periodic in the logarithm of spacetime scale with a period $\Delta$,
rather than continously scale-invariant. For both systems, the
critical solution seen by fine-tuning generic initial data has also
been obtained by a DSS ansatz, and similarly the critical exponent has
also been obtained from the linear perturbations of the resulting
exactly DSS (critical) solution. The massless scalar field has
$\Delta_{\rm scal}\simeq 3.44$ and $\gamma_{\rm scal}\simeq 0.37$
\cite{Choptuik1993, Gundlachscalar}, while the Yang-Mills (from now
on, YM) field has $\Delta_{\rm YM}\simeq 0.6$ and $\gamma_{\rm
  YM}\simeq 0.2$ \cite{ChoptuikChmajBizon, GundlachYM}.

Our system resembles the one recently considered by Kain \cite{Kain},
with the same YM field, but where the second matter field is a scalar
field transforming as a triplet under the $SU(2)$ symmetry. The main
difference is that those two matter fields are coupled directly
already on flat spacetime, as well as (inevitably) gravitationally. As
a consequence, while the triplet could consistently be switched off,
the YM field cannot\footnote{We note, however, that a scaling analysis
  shows that in an asymptotically self-similar solution the direct
  coupling becomes negligible in the limit of arbitrarily small
  scales.}. Kain finds approximately DSS critical solutions dominated
by one or the other field, but does not focus on the case in between
where both fields are in some sense equally strong.

Another closely related system is the most general ansatz for the
$SU(2)$ YM field compatible with spherical symmetry. This consists of
a ``sphaleronic'' degree of freedom in addition to the ``magnetic''
one considered in \cite{ChoptuikChmajBizon, GundlachYM} and in the
present paper, and so again there are two wave-like matter
fields. Critical collapse, including type II, in this system was
investigated numerically by Maliborski and Rinne
\cite{MaliborskiRinne}. The full system has a global $U(1)$ symmetry,
with a Mexican hat potential for the two fields. The magnetic field
$w$ and sphaleronic field $\omega$ can be reparameterised as
$w+i\omega=Se^{i\phi}$. The purely magnetic dynamics are then
recovered by setting $\phi$ to a constant. Our interpretation of the
results in Sec.~II.C of \cite{MaliborskiRinne} is that type II
critical collapse in the full system is dominated by the $S$ critical
solution (that is, the known critical solution for the purely magnetic
system), with the dynamics of $\phi$ subdominant.\footnote{We note
  also that a scaling analysis shows that the $S$ and $\phi$ dynamics
  remain coupled in the limit of increasingly small scales along an
  approximately self-similar solution.}

%%%%%%%%%%%%%%%%%%%%%%%%%%%%%%%%%%%%%%%%%%%%%%%%%%%%%%%%%%%%%%%%%%%%%%%%%%%%%%%

\section{Covariant form of the field equations}

%%%%%%%%%%%%%%%%%%%%%%%%%%%%%%%%%%%%%%%%%%%%%%%%%%%%%%%%%%%%%%%%%%%%%%%%%%%%%%%

Our first matter field is the minimally coupled scalar field $\psi$,
for which Choptuik \cite{Choptuik1993} originally discovered critical
phenomena. It obeys the 4-dimensional wave equation
\begin{equation}
\Box_4\psi:=\nabla_a\nabla^a\psi=0.  
\end{equation}
This wave equation holds on arbitrary spacetimes, but here we always
restrict to spherical symmetry.  (We use Latin letters for abstract
indices, and Greek ones for coordinate-specific ones, and
gravitational units such that $c=G=1$.)

Our second matter field is the purely magnetic hedgehog ansatz for an
$SU(2)$ Yang-Mills field in spherical symmetry, which was famously
considered for static solutions by Bartnik and Mckinnon \cite{BartnikMckinnon},
and later for critical collapse by Choptuik, Chmaj and Biz\'on
\cite{ChoptuikChmajBizon}. This can be parameterised by a single
spherically symmetric field $W$ as
\begin{eqnarray}
\label{Wansatz}
F&=&dW\wedge(\tau_1\,d\theta+\tau_2\,\sin\theta\,d\varphi) \nonumber
\\
&&-(1-W^2)\tau_3 d\theta\wedge\sin\theta\,d\varphi,
\end{eqnarray}
where $\tau_i$ are the Pauli matrices with $\tr
\tau_i\tau_j=\delta_{ij}$, and $(\theta,\varphi)$ are the usual angles
on the 2-spheres that are invariant under spherical symmetry. $W$ then
obeys an equation of motion that is the wave equation in the
2-dimensional spacetime reduced by the spherical symmetry, with a
potential term, namely
\begin{equation}
\label{boxW}
\Box_2 W:=R^2\nabla_a\left(R^{-2}\nabla^aW\right)=-{W(1-W^2)\over
  R^2}.
\end{equation}
Here the scalar $R$ is the area radius, so that the area of the
invariant 2-spheres is $4\pi R^2$. We stress that $W$ and its
equation of motion (\ref{boxW}) are defined only in a spherically
symmetric spacetime via the ansatz (\ref{Wansatz}).

In a static spacetime, there are two stable ground state solutions
$W=\pm 1$, while the third constant solution $W=0$ is unstable. Any
solution that is smooth at the origin must take the form $W=\pm
1+O(R^2)$. We pick the ground state $W=1$ and without loss of
generality write
\begin{equation} \label{chi}
W=1+R^2\chi,
\end{equation}
where now in any smooth solution $\chi$ is an even function of $R$
and generically $\chi=O(1)$ at the origin, as is also the case for the scalar
field $\psi$.

We write the spherically symmetric metric in relaxed notation as
\begin{equation}
g_{ab}=\textrm{diag}\left(\perp g_{ab},
R^2\gamma_{ab}\right),
\end{equation}
where $\gamma_{ab}$ is the unit metric on the 2-sphere and $\perp$
denotes the projection orthogonal to the invariant 2-spheres. The
Einstein equations are
\begin{equation}
G_{ab}=8\pi\left(T_{ab}^{(\psi)}+T_{ab}^{(W)}\right),
\end{equation}
where
\begin{eqnarray}
T_{ab}^{(\psi)}&=&\tilde T_{ab}^{(\psi)}-{1\over 2}g_{ab}\tilde
T^{(\psi)}, \\
\tilde T_{ab}^{(\psi)}&=& \nabla_a\psi\nabla_b\psi,
\end{eqnarray}
and
\begin{eqnarray}
\label{TabW}
T_{ab}^{(W)}&=&\tilde T_{ab}^{(W)}-{1\over 4}g_{ab}\tilde T^{(W)}, \\
\tilde T_{ab}^{(W)}&=& \tr F_{ab}F^{ab} \nonumber \\
&=& \textrm{diag}\left(\perp \tilde T_{ab},
PR^2\gamma_{ab}\right), \\
\perp \tilde T_{ab}&=&2R^{-2}\nabla_a W\nabla_b W, \\
\label{PYM}
P&=&R^{-2}\nabla_aW\nabla^aW+R^{-4}(1-W^2)^2.
\end{eqnarray}
The scalar $P$ can be interpreted as the tangential pressure in
spherical symmetry. $T_{ab}^{(\psi)}$ and $T_{ab}^{(W)}$ are conserved
separately, and these conservation laws imply the equations of
motion for $\psi$ and $W$.

%%%%%%%%%%%%%%%%%%%%%%%%%%%%%%%%%%%%%%%%%%%%%%%%%%%%%%%%%%%%%%%%%%%%%%%%%%%%%%%

\section{Double null coordinates}

%%%%%%%%%%%%%%%%%%%%%%%%%%%%%%%%%%%%%%%%%%%%%%%%%%%%%%%%%%%%%%%%%%%%%%%%%%%%%%%

We shall use double null coordinates, in terms of which the line
element becomes
\begin{equation}
\label{uvmetric}
ds^2=-2g\,R_{,v}\,du\,dv+R^2\left(d\theta^2+\sin^2\theta\,d\varphi^2\right).
\end{equation}
Here $g$ and $R$ are functions of $u$ and $v$ only. This means
  that surfaces of constant $u$ and $v$ are both null, and that the
  affinely parameterised null geodesics ruling the surfaces of
  constant $u$ have tangent vector $\nabla^a u$.  We fix the
remaining gauge freedom $u\to \tilde u(u)$, $v\to\tilde v(v)$ by
setting $R(0,v)=v/2$, and requiring that $u$ is proper time at the
regular centre $R=0$.

There are four algebraically independent components of the Einstein
equations. From these, we select one which is an ordinary differential
equation (from now on, ODE) for $g$ on the slices of constant $u$ and
another one which is a wave equation for $R$. The other two Einstein
equations are then redundant (they are replaced by suitable boundary
conditions at $R=0$). We also have wave equations for $\psi$ and
$\chi$. 

The four field equations we use can be arranged in the following hierarchy:
\begin{eqnarray}
\label{Dlng}
  {\cal D}(\ln g)&=&4\pi R\left[{\cal D}\psi)^2+2(R{\cal D}\chi
    +2\chi)^2\right], \\
\label{DRRu}
  {\cal D}(RR_{,u})&=&-{g\over2}+2\pi gR^2\chi^2(2+R^2\chi)^2,
  \\
\label{DRpsiu}
  {\cal D}(R\psi_{,u})&=&-R_{,u}{\cal D}\psi, \\
\label{DR2chiu}
  {\cal D}(R^2\chi_{,u})&=&-2RR_{,u}{\cal D}\chi
-4\pi gR^2\chi^3(2+R^2\chi)^2\nonumber \\
&&-{g\over2}R^2\chi^2(3+R^2\chi).
\end{eqnarray}
Here
\begin{equation}
{\cal D}f:={f_{,v}\over R_{,v}},
\end{equation}
so that ${\cal D}$ is the derivative with respect to $R$ along
  the null geodesics ruling the slices of
constant $u$. Eqs.~(\ref{Dlng}-\ref{DR2chiu}) can be solved for $g$,
$R_{,u}$, $\psi_{,u}$ and $\chi_{,u}$ in the above order by the
integration
\begin{equation}
{\cal I}f:=\int f R_{,v}\,dv
\end{equation}
along lines of constant $(u,\theta,\varphi)$, starting from
the centre. Because of factors of $R$, three of the startup
conditions are selected by regularity at $R=0$, and we fix the fourth
condition by imposing the gauge choice $g=1$ at $R=0$. This selection
and arrangement of equations resembles the form of the field equations
for the spherical scalar field of
\cite{GoldwirthPiran1987,GundlachPricePullin1994,Garfinkle1995} (but
with ${\cal D}$ and $\partial/\partial u$ applied to $\psi$ in the
opposite order), and also the scheme of \cite
{GomezPapadopoulosWinicour1994} for the vacuum Einstein equations on
null cones with a regular vertex (but in terms of double null
coordinates $u$ and $v$, rather than Bondi coordinates $u$ and $R$).

Using $dR=R_{,u}du+R_{,v}dv$, we can transform the metric (\ref{uvmetric})
from the double-null coordinates $(u,v)$ to the Bondi coordinates
$(u,R)$, obtaining
\begin{equation}
\label{uRmetric}
ds^2=-2g\,du\,dR-G\,du^2+R^2\left(d\theta^2+\sin^2\theta\,d\varphi^2\right),
\end{equation}
where
\begin{equation}
G:=-2gR_{,u}.
\end{equation}
At the centre $R=0$, we have $G=g^2$, and so, as already mentioned
above, we impose $g=1$ there in order to make $u$ proper time along
the worldline $R=0$. The quantity $\sqrt{G}\,du$ is proper time along
any curve of constant $(R,\theta,\varphi)$, and so $\sqrt{G}$ gives
the redshift of photons travelling from the central $R=0$ observer to
other constant $R$ and nonrotating observers, at constant retarded
time $u$. We can also define the compactness $2M/R$, where $M$ is the
Hawking mass, as
\begin{equation} \label{compactness}
{2M\over R}:=1-\nabla_aR\nabla^aR=1+2{R_{,u}\over g}.
\end{equation}

For the diagnosis of subcritical scaling we define the stress-energy
quantities 
\begin{eqnarray}
T_{(1)}&:=&{2\over g}\psi_u {\cal D}\psi, \\
T_{(2)}&:=&{2\sqrt 2\over
    g}\left(R\chi_{,u}+2R_{,u}\chi\right)\left(R{\cal
    D}\chi+2\chi\right), \\
T_{(3)}&:=&\chi^2(2+R^2\chi)^2.
\end{eqnarray}
These definitions are motivated by the identities ${T^a}_a^{(\psi)}=T_{(1)}$ and 
$T_{ab}^{(\psi)}T^{ab(\psi)}=T_{(1)}^2$ for the scalar field, 
and ${T^a}_a^{(\chi)}=0$ and
$T_{ab}^{(\chi)}T^{ab(\chi)}=T_{(2)}^2+T_{(3)}^2$ for the YM field. 
(The square of the total stress-energy is not a sum of
squares.)

%%%%%%%%%%%%%%%%%%%%%%%%%%%%%%%%%%%%%%%%%%%%%%%%%%%%%%%%%%%%%%%%%%%%%%%%%%%%%%%

\section{Numerical method}

%%%%%%%%%%%%%%%%%%%%%%%%%%%%%%%%%%%%%%%%%%%%%%%%%%%%%%%%%%%%%%%%%%%%%%%%%%%%%%%

Our numerical implementation follows that of \cite{Garfinkle1995} (for
similar, but not the same, field equations). We represent our fields
on a grid at fixed values of $v$, and numerically advance in the
retarded time $u$.

At every time step, we make a least-squares fit $\psi\simeq
\psi_0+\psi_1R+O(R^2)$, and $\chi\simeq \chi_0+\chi_1R+O(R^2)$ to the
four innermost grid points where $R>0$ (points where $R<0$ are no
longer evolved). We then substitute these expansions into the integral
expressions for $g$, $R_{,u}$, $\psi_{,u}$ and $\chi_{,u}$, obtaining
\begin{eqnarray}
g&=&1+2\pi R^2(\psi_1^2+8\chi_0^2)+O(R^3), \\
R_{,u}&=&-{1\over 2}-{\pi\over 3} \psi_1^2R^2+O(R^3), \\
\psi_{,u}&=&{1\over 2}\psi_1+O(R), \\
\chi_{,u}&=&{1\over 2}\chi_1+O(R).
\end{eqnarray}
The truncations indicated denote generically non-vanishing terms that
depend on the next order expansion coefficients. This expansion is
used for the first three grid points with $R>0$ to start up the
integration. For the remaining grid points, the integral ${\cal I}$ is
discretized with respect to $R$ by the trapezoid rule and ${\cal D}$
by symmetric finite differencing with respect to $R$ between adjacent
grid points, giving a scheme that is second-order accurate in
$v$. (The grid values of $v$ never appear explicitly, only those of
$R$ and the other fields). $R$, $\chi$ and $\psi$ are evolved in $u$
using {\blue the method of lines, and specifically} a second-order
Runge-Kutta integrator, with the expansions and integrals re-evaluated
for each Runge-Kutta substep. {\blue After the completion of each full
  time step we excise grid points with $R<0$, and continue evolving
  only those with $R>0$.\footnote{\blue We do not change the spatial
    grid within each time step in order for the differentiability
    assumptions underlying the method of lines to hold.}}

The scheme as described so far is strictly causal, {\blue in the sense
  that information does not travel numerically outside the continuum
  lightcones given by the lines of constant $u$ and $v$. Considering
  $u$ as the time coordinate, this means that} information does not
travel in the negative $v$ direction. Therefore no boundary condition
is required either mathematically or numerically at the outer boundary
$v=v_{\rm max}$. 

{\blue Stability of a finite difference algorithm for a hyperbolic
  problem usually requires a limit on the ratio of the time step to
  the space step, commonly called the Courant factor. A necessary
  upper bound on the Courant factor is given by the requirement of
  causality, that is, all physical characteristics must lie inside the
  numerical domain of influence of the finite difference stencil. The
  upper bound on the Courant factor sufficient for numerical stability
  is typically lower than the necessary bound from causality by some
  factor of order unity. In double null coordinates, by contrast, there is no
  upper bound on the Courant factor $\Delta u/\Delta v$ from
  causality, but of course some upper bound will still be required for
  numerical stability. A geometric condition involving only $\Delta
  u/\Delta v$ and the metric that can serve as an alternative to the
  causality condition is}
\begin{equation}
\label{Cdef}
|R_{,u}|\Delta u\le C R_{,v}\Delta v,
\end{equation}
{\blue where $C$ is a dimensionless factor of order unity (see also
  \cite {GundlachPricePullin1994}).}  We implement this as
\begin{equation}
\Delta u =C \min_i {2(R_{i}-R_{i-1})\over (R_{,u})_i+(R_{,u})_{i-1}}.
\end{equation}
This is evaluated before each full time step. {\blue We find
  empirically that the method is stable for test fields on Minkowski
  spacetime for $C\le 2.8$, but unstable for $C\ge 3.0$, for 100, 200,
  400 and 800 grid points, and so there seems to be a necessary and
  sufficient limit on $C$ approximately independent of resolution, as
  was intended with the condition (\ref{Cdef}). As one would expect
  for violating a Courant condition, for $C$ too large the code blows
  up rapidly and at grid spatial frequency. Allowing for a large
  safety margin in strongly curved spacetimes, we have used $C=0.1$
  for all our nonlinear simulations. Note that only the matter wave
  equations are subject to a Courant condition at all, as the Einstein
  equations we solve are hypersurface equations solved by integration
  on time slices (more precisely, along outgoing null rays).}

Although we use double-null coordinates, the fact that we have
parameterised $g_{uv}=-gR_{,v}$ implies that our time evolution stops
asymptotically as a marginally outer-trapped sphere is approached,
indicated in our coordinates by $R_{,v}=0$. We identify the formation
of a black hole by the time step going below a predefined threshold,
and we approximate the (Hawking) mass of the first outer-trapped
sphere by $R/2$ at the gridpoint where $R_{,v}$ is minimal.

Following Garfinkle, \cite{Garfinkle1995}, we use a very simple form
of mesh refinement adapted to the specific problem of critical
collapse in spherical symmetry. We identify, by hand, the approximate
value $v_0$ of $v$ corresponding to the point of largest curvature, or
the first outer-trapped sphere, in a suitable near-critical evolution
within a given one-parameter family of initial data. Once half of the
grid points with $v<v_0$ are excised because they have reached $R<0$,
we reuse the computer memory by fitting a new grid point in the middle
between each existing grid point. We assign values of $R$, $\psi$ and
$\chi$ to the new grid points by cubic Gauss interpolation, which for
this simple arrangement gives (in terms of the new grid index)
\begin{equation}
f_i={9\over 16}\left(f_{i+1}+f_{i-1}\right)-{1\over
  16}\left(f_{i+3}+f_{i-3}\right). 
\end{equation}
In this regridding step, {\blue information can propagate to negative
  $v$ by a distance of $(3/2)\Delta v_{\rm old}$ (the width of the
  interpolation stencil). It would be possible to avoid this by
  setting field values at the new grid points by extrapolation from
  the left (smaller $v$), but we would expect this to be noisy, and so
  have not attempted it. We note that the regridding does not happen after
  a fixed number of time steps, but at fixed moments of time
  independently of resolution, and so should not be thought of as part
  of the underlying time evolution scheme. We also note that the
  numerical scheme remains exactly causal in the presence of
  regridding, in the sense that the new outer boundary point is
  obtained in the same way as the new interior points, and no outer
  boundary condition is required.}

\begin{figure}
\includegraphics[scale=0.3, angle=0]{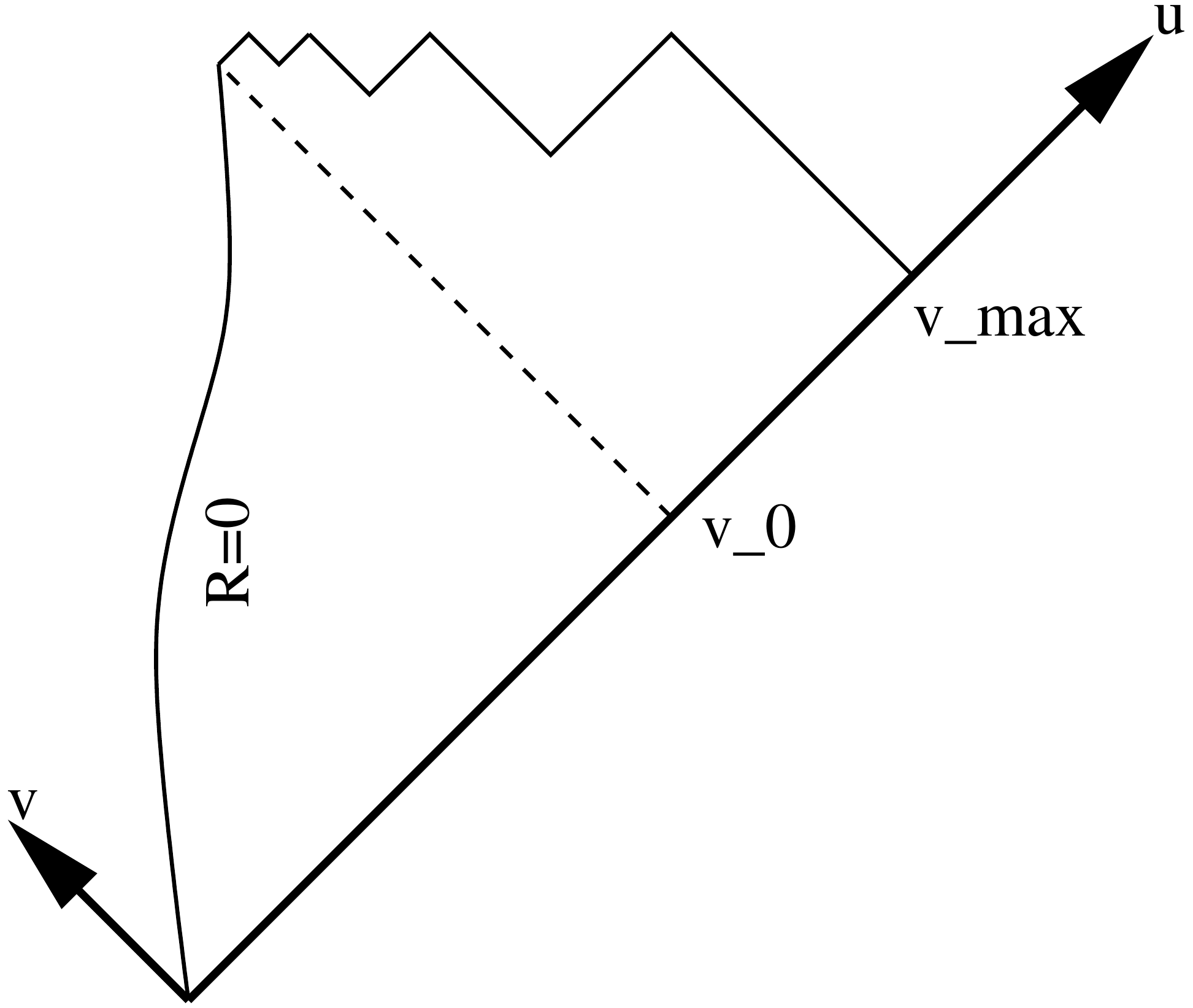} 
\caption{Schematic plot of the numerical domain in the $uv$-plane,
  focussing on the regridding process. For simplicity, this domain has
  only three regriddings. The regular centre $R=0$ is not
  in general at $u=v$.}
\label{fig:domain}
\end{figure}

In contrast to \cite{Garfinkle1995}, where
$v_0$ and $v_{\rm max}$ are identical, we have $v_{\rm max}>v_0$, and
we regrid by a factor of $2$ also in $v_0<v<v_{\rm max}$ after
discarding the outer half of it. The resulting numerical domain is
shown schematically in Fig.~\ref{fig:domain}. Having the buffer
outside of $v_0$ allows us to see what we would otherwise be missing
by potentially making $v_0$ too small, and seems to make the
regridding more robust, at the cost of only an insignificant increase
in computing time.

%%%%%%%%%%%%%%%%%%%%%%%%%%%%%%%%%%%%%%%%%%%%%%%%%%%%%%%%%%%%%%%%%
%%%%%%%%%%%%%%%%%%%%%%%%%%%%%%%%%%%%%%%%%%%%%%%%%%%%%%%%%%%%%%%%%%%%%%%%%%%%%%%

%%%%%%%%%%%%%%%%%%%%%%%%%%%%%%%%%%%%%%%%%%%%%%%%%%%%%%%%%%%%%%%%%%%%%%%%%%%%%%%

\section{Similarity solutions}

%%%%%%%%%%%%%%%%%%%%%%%%%%%%%%%%%%%%%%%%%%%%%%%%%%%%%%%%%%%%%%%%%%%%%%%%%%%%%%%

In any coordinates $x^\mu=(T,x,\theta,\varphi)$ adapted to spherical
symmetry and to DSS, a spherically symmetric spacetime is DSS if and
only if the metric takes the form
\begin{equation}
\label{DSSmetric}
g_{\mu\nu}=e^{-2T}\tilde g_{\mu\nu},
\end{equation}
where $\tilde g_{\mu\nu}$ is periodic in $T$ with some period
$\Delta$. In particular, the area radius $R$ must take the form
\begin{equation}
\label{tildeRdef}
R=e^{-T}\tilde R,
\end{equation}
with $\tilde R$ again periodic. 

A scalar-field $\psi$ whose stress-energy tensor is compatible with
this metric must itself be periodic in $T$ with the same period.

The equation of motion of $\chi$ (or $W$) is not compatible with exact
DSS, but there are solutions which are asymptotically DSS on small
scales as $T\to\infty$, and in which $\chi$ takes the form
\begin{equation} \label{chitilde}
\chi=e^T\tilde\chi
\end{equation}
with $\tilde\chi$ asymptotically periodic as $T\to\infty$. In
particular, the observed critical solution is of this form. In this
limit the two round brackets on the right-hand side of 
\begin{equation}
-{W(1-W^2)\over R^2}=\chi\left(1+e^{-T}\tilde
R^2\tilde\chi\right)
\left(2+e^{-T}\tilde R^2\tilde\chi\right)
\end{equation}
can be approximated by $1$ and $2$, respectively, as $T\to\infty$ in
(\ref{tildeRdef}), turning (\ref{boxW}) into
\begin{equation}
\label{boxchiscaleinvariant}
\Box_2 (R^2\chi)=2\chi,
\end{equation}
(In flat spacetime in the usual coordinates, where
$\Box_2=-\partial_t^2+\partial_r^2$, this equation is just
$-\chi_{,tt}+\chi_{,rr}+4r^{-1}\chi_{,r}=0$, with the $r^{-2}$ terms
on both sides canceling.)

One could consistently truncate the equation of motion (\ref{boxW})
for $W$ to (\ref{boxchiscaleinvariant}) and its conserved
stress-energy tensor (\ref{TabW}-\ref{PYM}) by replacing (\ref{PYM})
with
\begin{equation}
\label{Ptruncated}
P=R^{-2}\nabla_aW\nabla^aW+4\chi^2.
\end{equation}
We do not do this here, but rely on fine-tuning to make the solutions
asymptotically DSS on small scales.

Based on our null coordinates, we define the specific DSS-adapted
coordinates
\begin{equation}
\label{xTdef}
x:={R\over u_*-u}, \qquad T=-\ln(u_*-u),
\end{equation}
for a constant $u_*>0$ and $u<u_*$. (For $u>u_*$, both $x$ and $T$ are
undefined.) From the form (\ref{uRmetric}) of the metric it is clear
that the metric in coordinate $(x,T)$ is of the form
(\ref{DSSmetric}) and that the spacetime is discretely self-similar if
and only if $g$ and $G$ are periodic in $T$. Hence we will use $(x,T)$
as auxiliary coordinates in the diagnosis of discrete self-similarity.
In an (asymptotically) DSS solution, the metric quantity $2M/R$ is
also periodic, and so are the matter fields $\psi$ and
$\tilde\chi=e^{-T}\chi$. We define the stress-energy scalars $\tilde
T_{(i)}:=e^{-2T}T_{(i)}$ which are periodic in DSS, except that the
YM potential term $\tilde T_{(3)}=4\tilde\chi^2(1+O(e^{-T}))$ is only
asymptotically periodic on small scales.

%%%%%%%%%%%%%%%%%%%%%%%%%%%%%%%%%%%%%%%%%%%%%%%%%%%%%%%%%%%%%%%%%%%%%%%%%%%%%%%

\section{Numerical results}

%%%%%%%%%%%%%%%%%%%%%%%%%%%%%%%%%%%%%%%%%%%%%%%%%%%%%%%%%%%%%%%%%%%%%%%%%%%%%%%

Following some rough initial experimentation, we have used the
2-parameter family of initial data
\begin{eqnarray}
\psi(0,v)&=&0.075\,p(1-q)\, e^{-\left({R-0.75\over 0.25}\right)^2}, \\
\chi(0,v)&=&0.25\,pq\, e^{-\left({R-0.3\over 0.1}\right)^2}.
\end{eqnarray}
Recall also that we make the gauge choice $R(0,v)=v/2$.  
Here $q=0$ corresponds to pure scalar field data and $q=1$ to pure YM
data. We have adjusted the other parameters so that for both $q=0$ and
$q=1$ the critical value of $p$ is $p_*\simeq 1$, and they have
similar values of $v_0\simeq
3.4$. We use $v_{\rm max}=4$ (so that $0\le R\le 2$ initially) and 401
grid points throughout.

We find that, for a given value of $q$, it is sufficient to carry out
about 30 bisections without regridding in order to get a good enough
estimate of $v_0$. We then fix $v_0$, (or rather, the corresponding
grid point index $i_0$) and carry out 50 bisections from scratch with
up to 8 regriddings in each evolution. (The buffer $v_0<v<v_{\rm max}$
allows us to do this without losing the accumulation point from the
grid and without the need for the iterative refinement of $v_0$
described in \cite{Garfinkle1995}). This simple process gives us mass
and curvature scaling down to the machine precision limit
$|p-p_*|\simeq 10^{-15}$. It is remarkable that this takes only a few
minutes of computation on a laptop, with negligible memory.

How many scale echos we see in our most fine-tuned evolutions with
$|p-p_*|\simeq 10^{-15}$ depends on both $\Delta$ and $\gamma$.  For
$q=0$ (pure scalar), we expect to see DSS over a range $\Delta T\simeq 15\cdot
\ln 10\cdot \gamma_{\rm scal}\simeq 12.8$, corresponding to $\Delta
T/\Delta_{\rm scal}\simeq 3.5$ scale oscillations, and this is what we
observe. Similarly, for $q=1$ (pure
YM), we get $\Delta T\simeq 15\cdot \ln 10\cdot \gamma_{\rm YM}\simeq
6.9$, corresponding to $\Delta T/\Delta_{\rm YM}\simeq 11.5$ scale
oscillations.

\begin{figure}
\includegraphics[scale=0.7, angle=0]{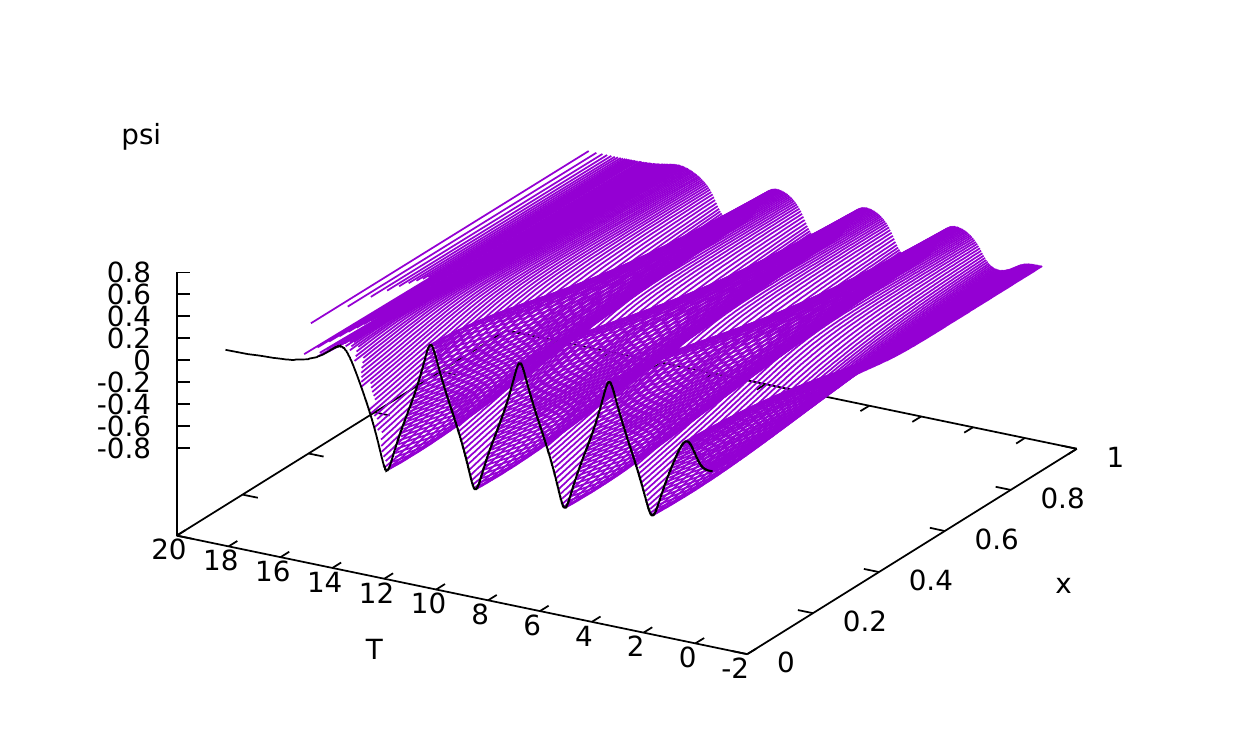} 
\caption{The scalar field $\psi(x,T)$ for optimal fine-tuning with
  $q=0$ (only the scalar field is present). The regular centre $x=0$
  runs along the front edge of the graph, with the values of $\psi$
  there emphasised by a black line line.}
\label{fig:psi00}
\end{figure}

\begin{figure}
\includegraphics[scale=0.7, angle=0]{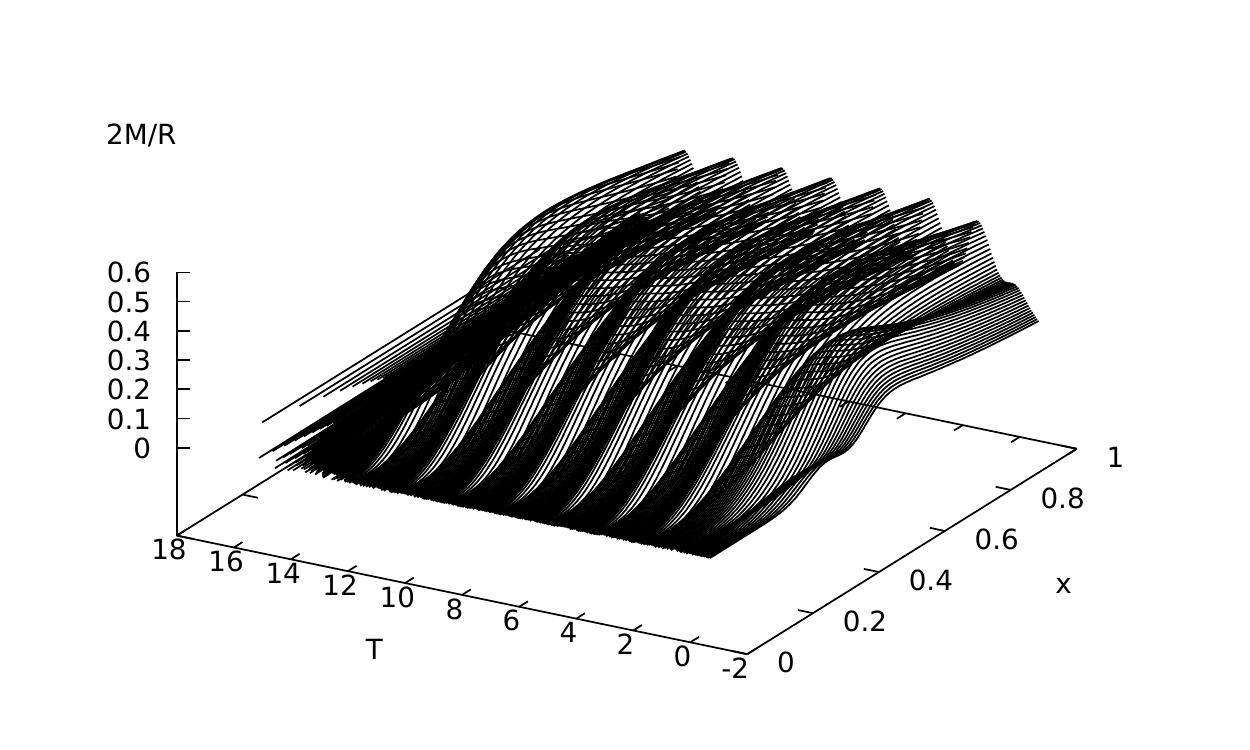} 
\caption{The compactness $(2M/R)(x,T)$, defined in Eq.~(\ref{compactness}),
  for the same solution as in Fig.~\ref{fig:psi00}.}
\label{fig:2MoR00}
\end{figure}

\begin{figure}
\includegraphics[scale=0.6, angle=0]{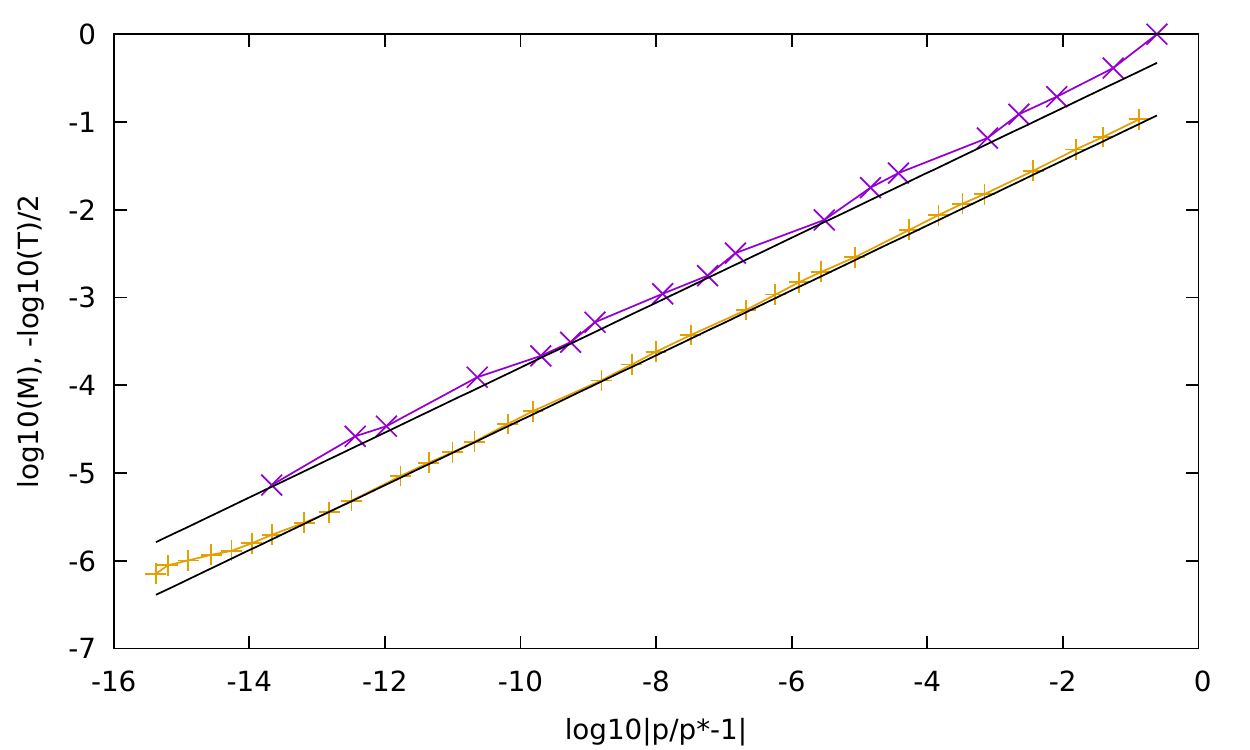} 
\caption{Mass and curvature scaling laws for the pure scalar
  field. The upper curve (purple) represents $-(1/2)\log_{10}({\rm
    max}T_{(1)})$, and the lower curve (orange) represents
  $\log_{10}M$, both against $\log_{10}|p/p_*-1|$. The straight grey
  lines corresponding to $\gamma=0.37$ are shown for comparison. }
\label{fig:scalinglaws_0}
\end{figure}

\begin{figure}
\includegraphics[scale=0.7, angle=0]{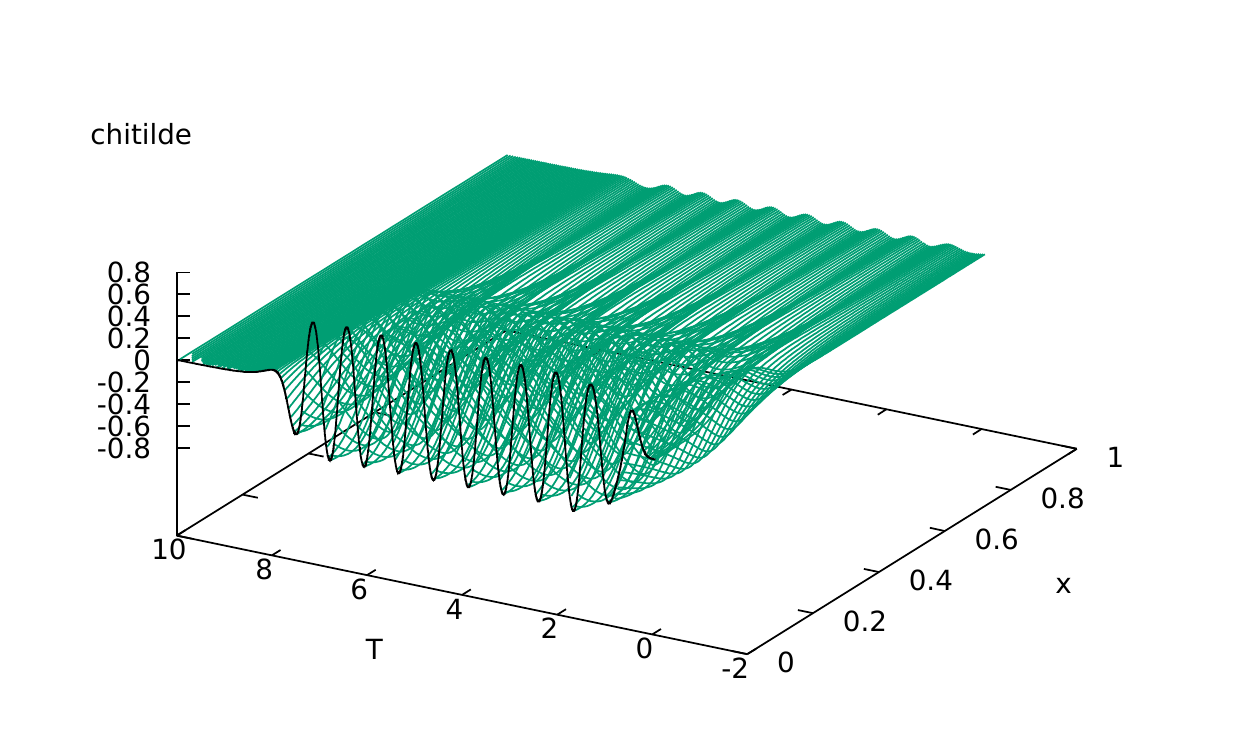} 
\caption{The rescaled Yang-Mills function $\tilde\chi(x,T)$ [see
    Eqs.~(\ref{chi}) and (\ref{chitilde})] for optimal
  fine-tuning with $q=1$ (only the YM field is present). Note the
  range of $T$ here ($-2<T<10$) is not the same as in
  Fig.~\ref{fig:psi00} ($-2<T<20$). Again the values of $\tilde\chi$
  at the regular centre $x=0$ are emphasised by a black line line.}
\label{fig:chi10}
\end{figure}

\begin{figure}
\includegraphics[scale=0.7, angle=0]{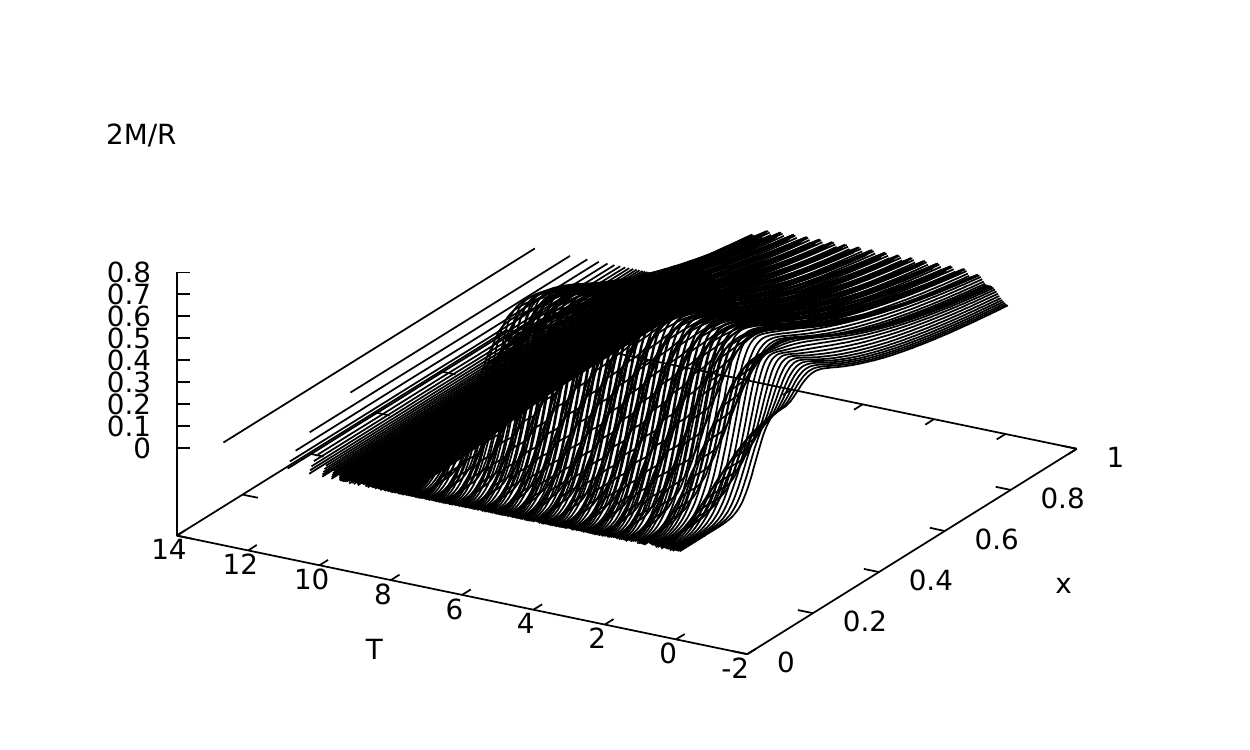} 
\caption{The compactness $(2M/R)(x,T)$ for the same solution as in
  Fig.~\ref{fig:chi10}.}
\label{fig:2MoR10}
\end{figure}

\begin{figure}
\includegraphics[scale=0.6, angle=0]{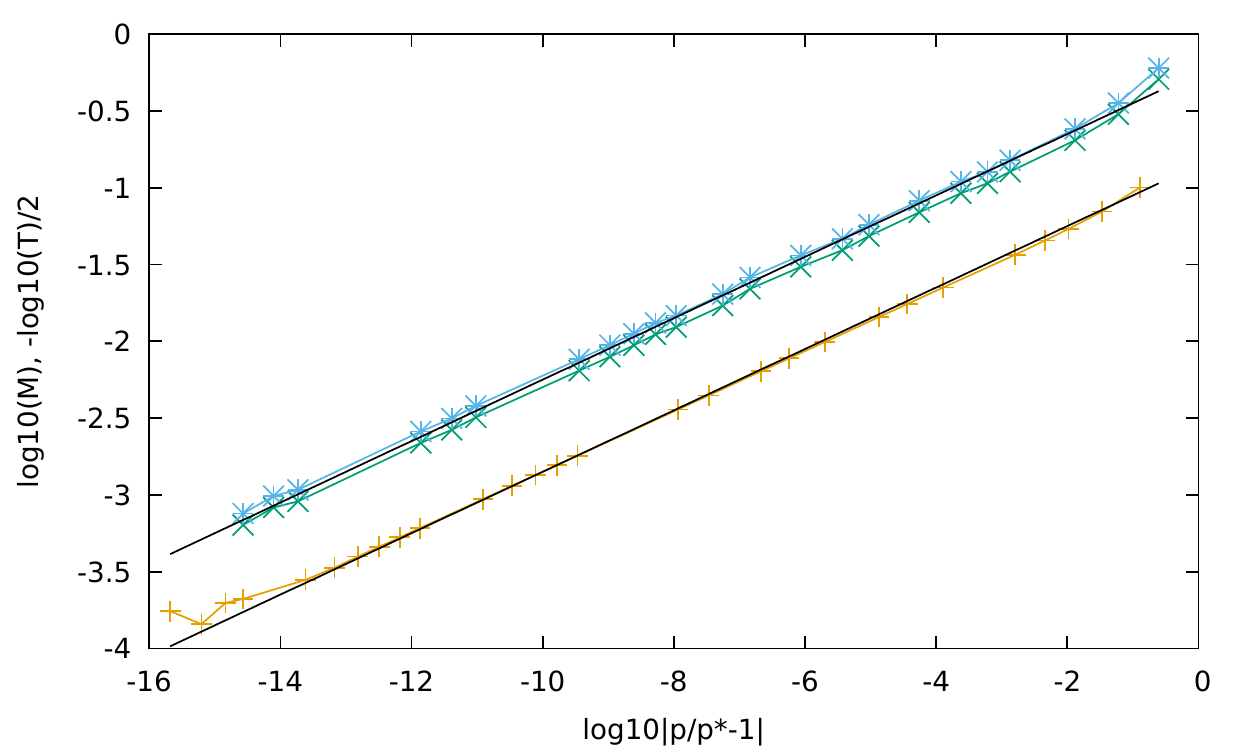} 
\caption{Mass and curvature scaling laws for the pure YM field, as
  in Fig.~\ref{fig:scalinglaws_0} except that the two upper curves now
  show $T_{(2,3)}$ (green and blue). Grey lines $\gamma=0.20$ shown
  for comparison. }
\label{fig:scalinglaws_1}
\end{figure}

We initially determine $u_*$ as the approximate value of retarded time
for black hole formation or largest curvature. We then plot $2M/R$,
$\psi$, $\tilde\chi$ and the $\tilde T_{(i)}$ against the resulting
$x$ and $T$, and further adjust $u_*$ to make these as periodic in $T$
as possible. (The $\tilde T_{(i)}$ are the most sensitive tool for
this.)  As a test of the correctness of our methods, we clearly see
the expected number of oscillations at the correct periods, and the
expected scaling of the black hole mass and the curvature terms
$T_{(i)}$ with the correct critical exponents, as well as the periodic
``wiggle'' in the curvature scaling laws that comes from DSS
\cite{Gundlachscalar,HodPiran1995}.  The scalar field $\psi$ in the
pure scalar field critical solution is shown in Fig.~\ref{fig:psi00},
and the compactness $2M/R$ of the corresponding spacetime in
Fig.~\ref{fig:2MoR00}. The rescaled YM field $\tilde\chi$ in the pure
YM critical solution is shown in Fig.~\ref{fig:chi10}, and the
compactness $2M/R$ of the corresponding spacetime in
Fig.~\ref{fig:2MoR10}. The corresponding scaling laws are shown in
Figs~\ref{fig:scalinglaws_0} and \ref{fig:scalinglaws_1},
respectively.  (We plot $\ln M$ and $-1/2\ln T_{(i)}$ against
$\ln|p/p*-1|$.)

%%%%%%%%%%%%%%%%%%%%%%%%%%%%%%%%%%%%%%%%%%%%%%%%%%%%%%%%%%%%%%%%%
%%%%%%%%%%%%%%%%%%%%%%%%%%%%%%%%%%%%%%%%%%%%%%%%%%%%%%%%%%%%%%%%%

%%%%%%%%%%%%%%%%%%%%%%%%%%%%%%%%%%%%%%%%%%%%%%%%%%%%%%%%%%%%%%%%%
%%%%%%%%%%%%%%%%%%%%%%%%%%%%%%%%%%%%%%%%%%%%%%%%%%%%%%%%%%%%%%%%%

%%%%%%%%%%%%%%%%%%%%%%%%%%%%%%%%%%%%%%%%%%%%%%%%%%%%%%%%%%%%%%%%%
%%%%%%%%%%%%%%%%%%%%%%%%%%%%%%%%%%%%%%%%%%%%%%%%%%%%%%%%%%%%%%%%%

%%%%%%%%%%%%%%%%%%%%%%%%%%%%%%%%%%%%%%%%%%%%%%%%%%%%%%%%%%%%%%%%%
%%%%%%%%%%%%%%%%%%%%%%%%%%%%%%%%%%%%%%%%%%%%%%%%%%%%%%%%%%%%%%%%%

%%%%%%%%%%%%%%%%%%%%%%%%%%%%%%%%%%%%%%%%%%%%%%%%%%%%%%%%%%%%%%%%%
%%%%%%%%%%%%%%%%%%%%%%%%%%%%%%%%%%%%%%%%%%%%%%%%%%%%%%%%%%%%%%%%%

%%%%%%%%%%%%%%%%%%%%%%%%%%%%%%%%%%%%%%%%%%%%%%%%%%%%%%%%%%%%%%%%%
%%%%%%%%%%%%%%%%%%%%%%%%%%%%%%%%%%%%%%%%%%%%%%%%%%%%%%%%%%%%%%%%%

We next add a YM test field to the scalar field critical solution, and
a scalar test field to the YM critical solution. (In practice, we
evolve $q=\epsilon$ and $q=1-\epsilon$ with $\epsilon=10^{-12}$. This
still requires a small adjustment to $p_*$, but makes no visible
difference to the dominant field and metric, while the perturbation
field obeys an essentially linear equation of motion.)

We can formally write the purely YM critical solution as $Z_{*\rm
  YM}(x,T)$, which obeys
\begin{equation}
Z_{*\rm YM}(x,T+\Delta_{\rm YM})=Z_{*\rm YM}(x,T), 
\end{equation}
for the dependent variables $Z=(\tilde\chi,g,G,R)$. Separation of
variables then allows us to consistently look for solutions of the
scalar test field equation of the form
\begin{equation}
\label{testfieldmode}
\psi(x,T) = e^{\lambda T}\psi_\lambda(x,T), 
\end{equation}
where $\psi_\lambda(x,T)$ is a complex mode function, with the same
period $\Delta_{\rm YM}$ in $T$ as the background solution.  The
complex frequency $\lambda$ arises from solving a boundary value
problem, imposing periodicity of $\psi_\lambda(x,T)$ in $T$ and
regularity at the regular centre and the past light cone of the
singularity of the background spacetime. We expect a discrete set of
complex conjugate pairs of such modes to exist. From a complex mode, a
specific real solution is then obtained as
\begin{equation}
\label{testfieldmodebis}
\psi(x,T) = \Re\left[ e^{i\delta} e^{\lambda T}\psi_\lambda(x,T)\right], 
\end{equation}
for an arbitrary value of the phase $\delta$. Similar expressions hold
for a YM test field $\tilde\chi$ on the scalar field critical solution
background with variables $Z=(\psi,g,G,R)$, and period
$\Delta_{\rm scal}$.

A time evolution with generic initial data for the linear test field
is dominated with increasing $T$ by the mode or complex conjugate pair
of modes with the largest real part of $\lambda$. Our numerical
results are compatible with this theoretical result, with the dominant
$\lambda=-0.40+3.7i$ for the YM field on the scalar field background,
and $\lambda=0.09+4.2i$ for the scalar field on the YM background. To
our surprise, the YM perturbation on the scalar field background
decays but the scalar field perturbation on the YM background grows
exponentially in $T$.

Next we explore the black-hole threshold in our two-parameter set of
initial data by fixing $q$ and fine-tuning $p$ to its ($q$-dependent)
critical value. For all values of $q$, we find a single critical
solution that is ``shared'' by the two matter fields.  There is
complementary evidence for this from the scaling laws (varying $p$) on
the one hand and from single ``best'' near-critical time evolutions
(with a particular $|p/p*-1|\sim 10^{-15}$) on the other.

The evidence from varying $p$ is that the black hole mass and the
maximum values of the energy densities $T_{(i)}$ achieved in a given
evolution show clear approximate power-law scaling with the same
$p_*$. By ``approximate'' power-law scaling we mean that the exponent
$\gamma$ for each fixed $q$ depends weakly on $\ln|p_*-p|$. In a plot
of $\ln M$ versus $\ln|p-p_*|$, this appears as an almost straight
line. For $q=0.9$, $0.92$ and $0.93$, we see a break in the slope,
with the larger slope (larger $\gamma$) closer to the black-hole
threshold.

\begin{figure}
\includegraphics[scale=0.6, angle=0]{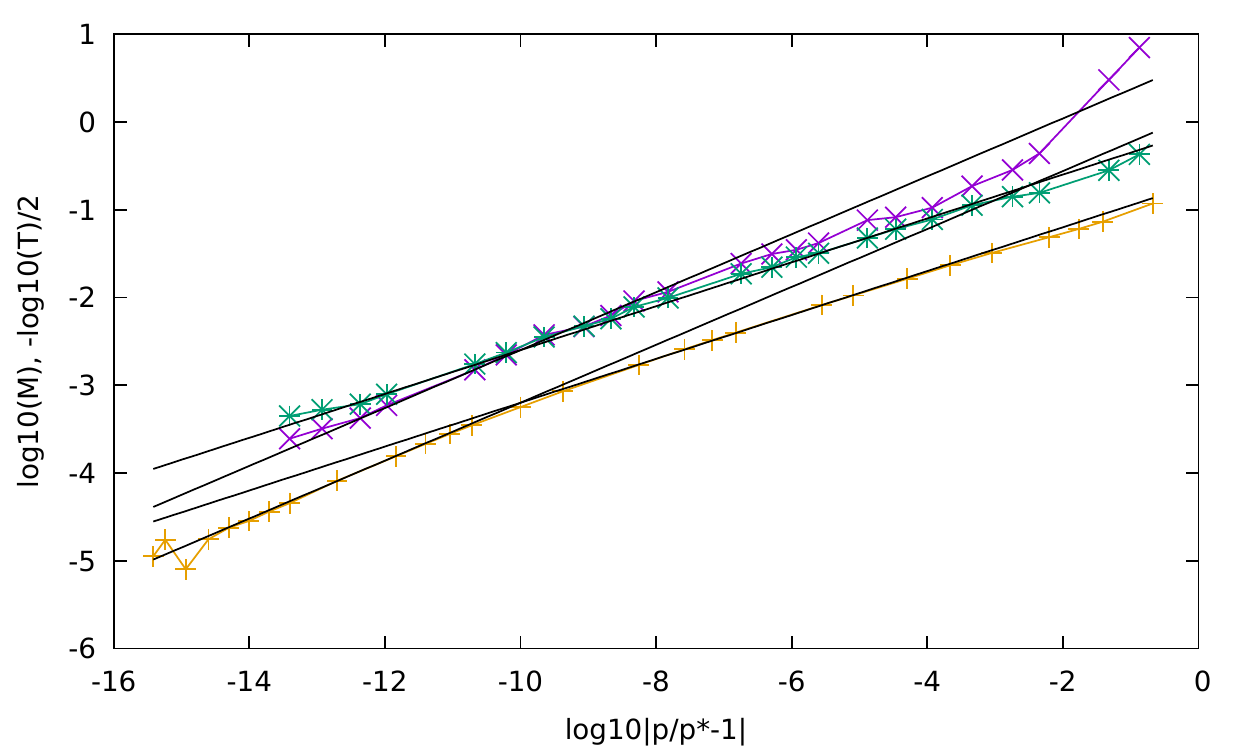} 
\caption{Mass and curvature scaling laws for the mixed initial data
  with $q=0.92$. As in Fig.~\ref{fig:scalinglaws_0} except
  that the two upper curves now show $T_{(1,2)}$ (purple and green).
  $T_{(3)}$ is not shown for clarity, but is close to $T_{(2)}$, as it
  was in Fig.~\ref{fig:scalinglaws_1}. There are now two pairs of
  straight grey lines for comparison, corresponding to $\gamma=0.25$
  and $\gamma=0.33$. These illustrate a break in the scaling laws
  around $|p/p*-1|=10^{-10}$.}
\label{fig:scalinglaws_092}
\end{figure}

Fig.~\ref{fig:scalinglaws_092} demonstrates this for $q=0.92$, where
the break is roughly in the middle of our fine-tuning range:
at low fine-tuning (on the right), the YM field dominates the
stress-energy (recall we are plotting {\em minus} the log of
$T_{(i)}$), and everything scales with $\gamma=0.25$, while at high
fine-tuning (on the left), the scalar field dominates, and everything
scales with $\gamma=0.33$. For $q=0.9$, the critical exponent breaks
from $0.25$ to $0.37$, and for $q=0.93$ from $0.23$ to $0.27$. (These
are all rough values fitted by eye, as illustrated in
Fig.~\ref{fig:scalinglaws_092}). By contrast, for $q=0.95$ we see a
constant critical exponent $0.22$, with the YM field dominant
throughout our fine-tuning range. Compare these values with
$0.37$ for the pure scalar field, $q=0$, and $0.20$ for the pure YM
field, $q=1$.

%%%%%%%%%%%%%%%%%%%%%%%%%%%%%%%%%%%%%%%%%%%%%%%%%%%%%%%%%%%%%%%%%
%%%%%%%%%%%%%%%%%%%%%%%%%%%%%%%%%%%%%%%%%%%%%%%%%%%%%%%%%%%%%%%%%

%%%%%%%%%%%%%%%%%%%%%%%%%%%%%%%%%%%%%%%%%%%%%%%%%%%%%%%%%%%%%%%%%
\begin{figure}
\includegraphics[scale=0.7, angle=0]{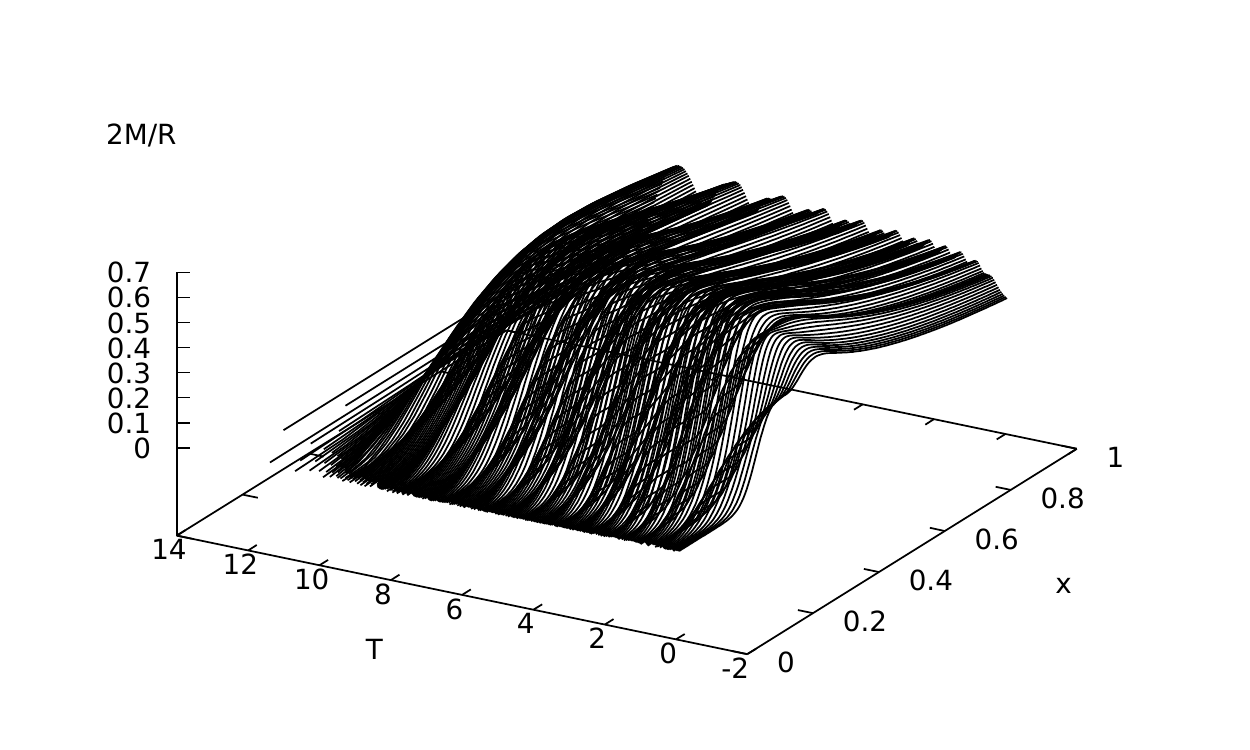} 
\caption{The compactness $2M/R$ against $x$ and $T$, in our ``best''
  subcritical time evolution with $q=0.92$.}
\label{fig:2MoR_092}
\end{figure}
%%%%%%%%%%%%%%%%%%%%%%%%%%%%%%%%%%%%%%%%%%%%%%%%%%%%%%%%%%%%%%%%%

%%%%%%%%%%%%%%%%%%%%%%%%%%%%%%%%%%%%%%%%%%%%%%%%%%%%%%%%%%%%%%%%%
\begin{figure}
\includegraphics[scale=0.6, angle=0]{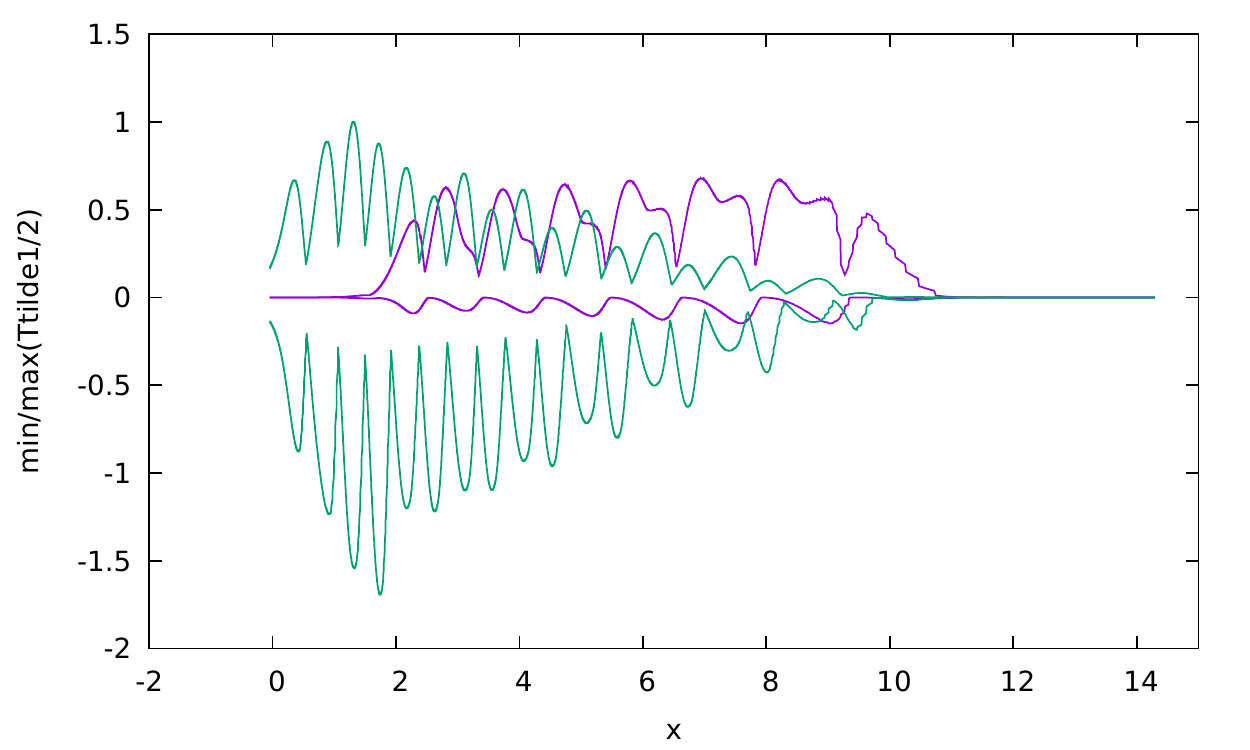} 
\caption{$\min_x$ and $\max_x$ of $T_{(1)}$, (scalar field
  stress-energy, purple) and similarly for $T_{(2)}$ (YM kinetic
  stress-energy, green), against $T$, in our ``best'' subcritical time
  evolution with $q=0.92$. We see that for small $T$ this spacetime is
dominated by the YM stress-energy, but by the scalar field
stress-energy at large $T$.}
\label{fig:Tminmax_092}
\end{figure}
%%%%%%%%%%%%%%%%%%%%%%%%%%%%%%%%%%%%%%%%%%%%%%%%%%%%%%%%%%%%%%%%%

%%%%%%%%%%%%%%%%%%%%%%%%%%%%%%%%%%%%%%%%%%%%%%%%%%%%%%%%%%%%%%%%%
\begin{figure}
\includegraphics[scale=0.6, angle=0]{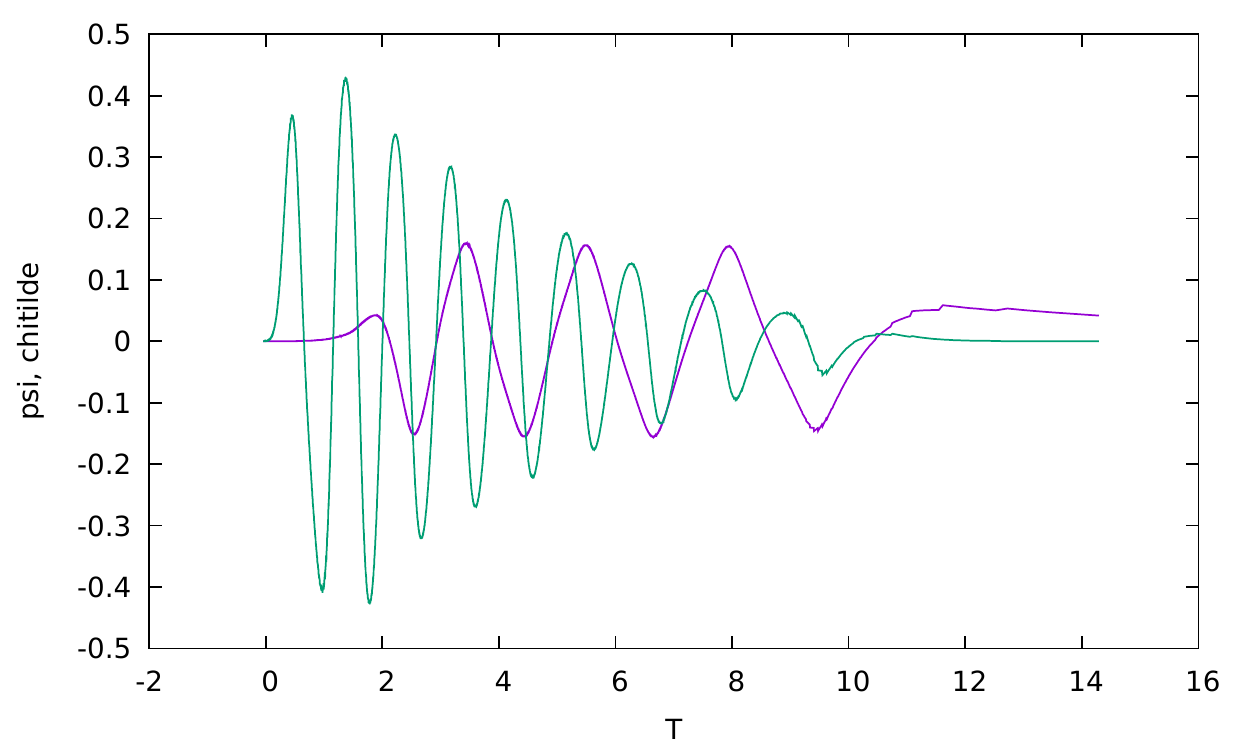} 
\caption{$\psi$ (scalar field, purple) and $\tilde\chi$ (rescaled YM
  field, green) at the origin, against $T$, in our most closely fine-tuned
  subcritical time evolution with $q=0.92$. Consistently with
  Fig.~\ref{fig:Tminmax_092}, the YM field decays and the scalar field
  takes over. Compare these curves with the periodic values at the
  origin $x=0$ in the two single-field critical solutions, shown as
  the black lines in
  Figs.~\ref{fig:psi00} and~\ref{fig:chi10}.}
\label{fig:psichi_092}
\end{figure}
%%%%%%%%%%%%%%%%%%%%%%%%%%%%%%%%%%%%%%%%%%%%%%%%%%%%%%%%%%%%%%%%%

The evidence from individual near-critical time evolutions is more
complicated, and is summarised in
Figs.~\ref{fig:2MoR_092}-\ref{fig:psichi_092}. The key observation is
that we can always fit the coordinates $x$ and $T$, and the rescalings
of $\chi$ to $\tilde\chi$ and the $T_{(i)}$ to $\tilde T_{(i)}$, with
a single choice of the parameter $u_*$ for all these variables and for
both matter fields. The two matter fields share the same accumulation
point of scale echos, and their stress-energies scale with the same
factor $(u_*-u)^{-2}$.

However, all quantities that would be strictly periodic in one of the
two single-field critical solutions are far from periodic in
mixed-field near-critical evolutions. The spacetime observable
$(2M/R)(x,T)$ switches over gradually with increasing $T$ from the
structure familiar from the pure YM solution (a maximum over $x$ of
around $0.7$, with small-amplitude high-frequency modulation in $T$)
to that of scalar field solution (a maximum over $x$ of around $0.5$,
with large low-frequency modulation in $T$). This is illustrated for
$q=0.92$ in Fig.~\ref{fig:2MoR_092}.

Moreover, $\psi$ and its stress-energy $\tilde T_{(1)}$ oscillate
(in a quasiperiodic way) and grow in $T$, while $\tilde\chi$
and its stress-energy $\tilde T_{(2,3)}$ oscillate and
decay. Fig.~\ref{fig:Tminmax_092} illustrates this switchover from
$\tilde T_{(2,3)}$ to $\tilde T_{(1)}$ during the evolution of
near-subcritical initial data with $q=0.92$.
Fig.~\ref{fig:psichi_092} illustrates the fields $\psi$ and
$\tilde\chi$ generating these stress-energy sectors. These should be
compared against Figs.~\ref{fig:psi00} and \ref{fig:chi10}.

For clarity, when we show both fields together we do not plot against
$x$ and $T$, but only the maxima and minima of $\tilde T_{(1,2)}$ over
$x$ [in effect, a sideways view of the corresponding $(x,T)$ plot],
and the central values of $\psi$ and $\tilde\chi$, all against $T$.

%%%%%%%%%%%%%%%%%%%%%%%%%%%%%%%%%%%%%%%%%%%%%%%%%%%%%%%%%%%%%%%%%%%%%%%%%%%%%%%

\section{Conclusions}

%%%%%%%%%%%%%%%%%%%%%%%%%%%%%%%%%%%%%%%%%%%%%%%%%%%%%%%%%%%%%%%%%%%%%%%%%%%%%%%

The numerical limits of fine-tuning do not allow us to follow the
putative critical solution for given $q$ down to arbitrarily large $T$, but
our observations are consistent with the assumption that in the limit
of perfect fine-tuning our time evolutions for all $0<q<1$ are heading
towards a critical solution that ends in a curvature singularity as
$T\to\infty$, with the scalar field increasingly dominant as $T$ increases.

Moreover, we conjecture that there is a family of fully nonlinear
solutions that start from the pure YM critical solution at
$T\to-\infty$ and end at the pure scalar field critical solution at
$T\to\infty$, that show the scaling of the metric and matter fields
with respect to an accumulation point $u_*$ typical of similarity
solutions, but where the spacetime is scale-periodic with a definite
period (DSS) only in the limits $T\to\pm\infty$. (This period is
$\Delta_{\rm YM}$ as $T\to-\infty$ and the YM field dominates, and
$\Delta_{\rm scal}$ as $T\to\infty$ and the scalar field
dominates.)\footnote{An exactly DSS solution for the YM field exists
  only in the limit $T\to\infty$ where $\tilde\chi$ varies over
  spacetime scales much smaller than the length scale set by the
  dimensionful YM self-coupling constant. Hence if we talk about the
  YM critical solution as $T\to-\infty$ (large scales), we really mean
  the critical solution for the simplified matter model
  (\ref{boxchiscaleinvariant},\ref{Ptruncated}) in which the YM field
  self-coupling is ignored.} In between they are at best
quasiperiodic, in the sense that we expect a Fourier transform in $T$
to show broad frequency peaks around multiples of both
$2\pi/\Delta_{\rm scal}$ and $2\pi/\Delta_{\rm YM}$. We propose the
term quasi-periodically self-similar (QSS) for such solutions. (In our
numerical experiments, we have not obtained this solution for a large
enough range of $T$ to usefully take such a Fourier transform.
Rather, we appeal to Fig.~\ref{fig:2MoR_092} to illustrate what we
mean by QSS, in contrast to Figs.~\ref{fig:2MoR00} and
\ref{fig:2MoR10}, which show approximately DSS spacetimes.)

We can be more specific about the size of this family of solutions if
we assume that at the YM starting point, where the scalar field is a
test field, this test field can be approximated by the most rapidly
growing (largest $\Re\lambda>0$) mode of the form
(\ref{testfieldmode}). There is then only a one-parameter family of
such QSS solutions, parameterised by the constant phase $\delta$ in
(\ref{testfieldmodebis}). Modulo this assumption, we conjecture that
the critical solutions we find at different $q$ by fine-tuning $p$ are
sections of this one-parameter family of QSS solutions, for starting
values of $T$ and values of $\delta$ that depend on $q$. (The range of
$T$ seen depends on the degree of fine-tuning we can achieve).

%%%%%%%%%%%%%%%%%%%%%%%%%%%%%%%%%%%%%%%%%%%%%%%%%%%%%%%%%%%%%%%%%
\begin{figure}
\includegraphics[scale=0.27, angle=0]{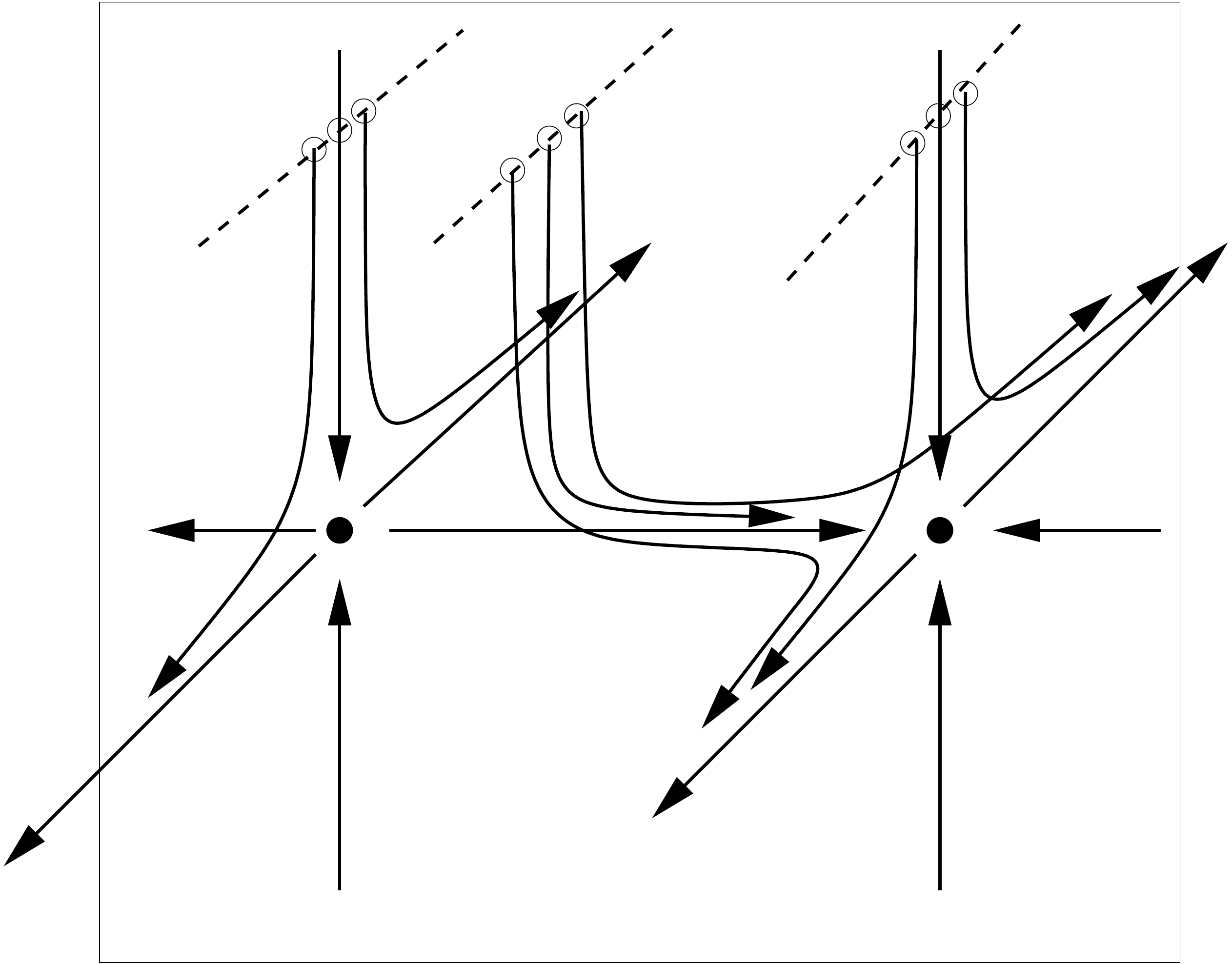} 
\caption{Schematic conjectured phase space picture, with the
  infinite-dimensional phase space represented in three
  dimensions. The framed plane represents the black hole threshold (in
  reality a hypersurface). All arrow lines represent trajectories
  (spacetimes). The filled dots represent fixed points (DSS
  spacetimes): the YM critical solution, with two unstable modes, on
  the left, and the scalar critical solution, with one unstable mode,
  on the right. An infinite number of phase space dimensions of the
  black hole threshold are suppressed, and with them an infinite
  number of stable modes of each fixed point within the black hole
  threshold. The horizontal straight arrow line linking the two
  represents the QSS solution conjectured in the Conclusions. The
  three dashed lines represent three families of initial data with
  $q=0$ (left), $q=1$ (right) and an intermediate value of $q$
  (middle). Hollow dots represent initial data with $p<p_*$, $p=p_*$
  and $p>p_*$ for each family.}
\label{fig:phasespace}
\end{figure}
%%%%%%%%%%%%%%%%%%%%%%%%%%%%%%%%%%%%%%%%%%%%%%%%%%%%%%%%%%%%%%%%%

%%%%%%%%%%%%%%%%%%%%%%%%%%%%%%%%%%%%%%%%%%%%%%%%%%%%%%%%%%%%%%%%%
\begin{figure}
\includegraphics[scale=0.27, angle=0]{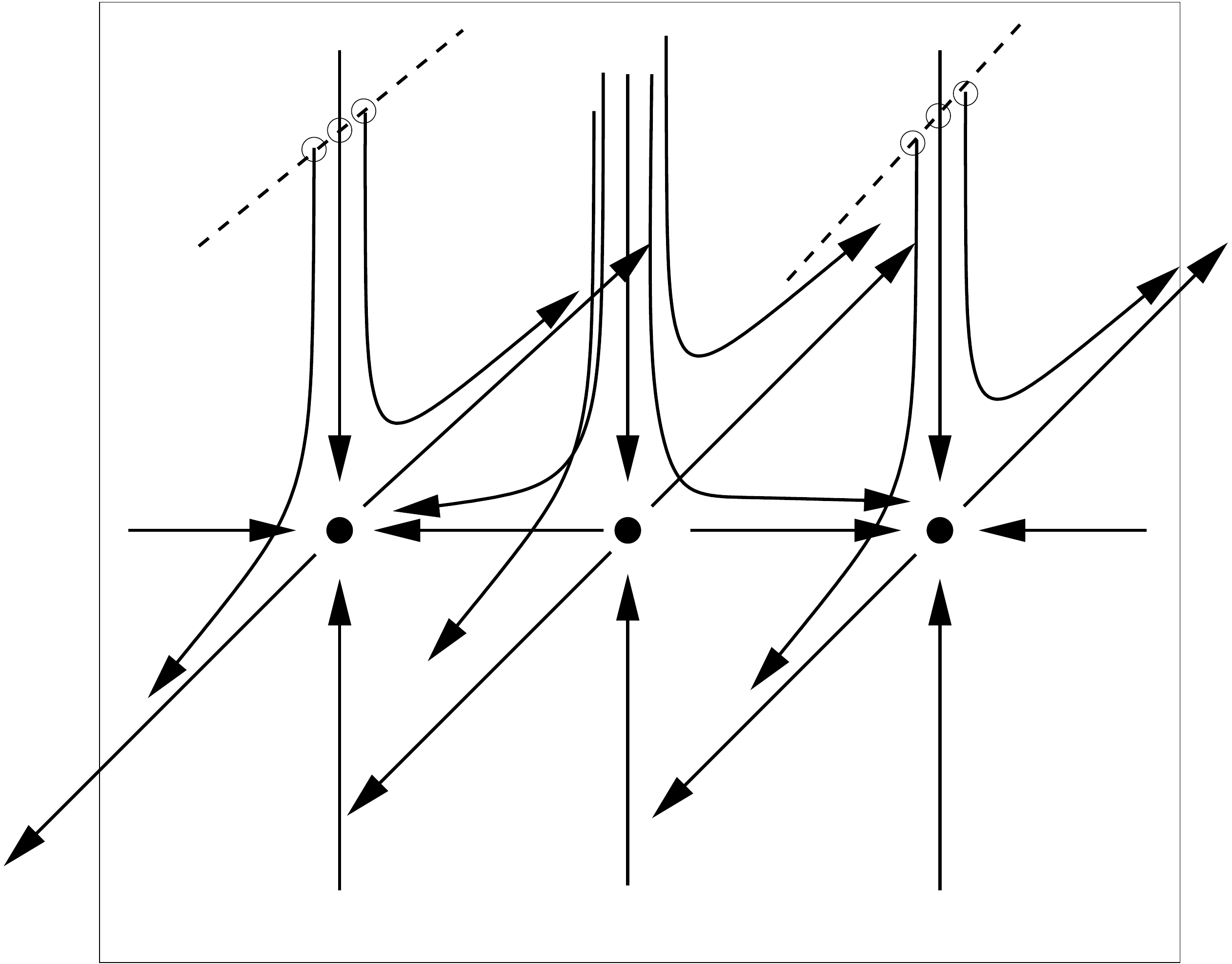} 
\caption{An alternative hypothetical phase space picture. Here the
  middle fixed point has two unstable modes, while the left and right
  ones have one each. This picture is definitely {\em not} realized
  for the spherical Einstein-YM-scalar system investigated in this
  paper, but might be realised for another system with two competing
  massless fields.}
\label{fig:phasespace2}
\end{figure}
%%%%%%%%%%%%%%%%%%%%%%%%%%%%%%%%%%%%%%%%%%%%%%%%%%%%%%%%%%%%%%%%%

Our main evidence for this conjecture is that at small $q$ the YM
field dominates down to the best fine-tuning we can manage (but the
subdominant scalar field grows exponentially), that at large $q$ the
scalar field dominates from the beginning of the DSS phase (and the YM
field continues to decay exponentially), while in an intermediate
range of $0.90\lesssim q\lesssim 0.93$ the scalar field takes over
within the observed range of $T$, at larger $T$ for smaller $q$.

This model can be summarised in the schematic phase space picture of
Fig.~\ref{fig:phasespace}. Here, any point in the phase space is an
initial data set, up to an overall length scale, parameterised in our
case as $(\psi(x),\tilde\chi(x))$, and a time evolution curve
corresponds to a spacetime, in our case in null slicing, again up to
an overall scale, with the time $T$ of the dynamical system
determining the missing scale as $e^{-T}$. In this picture, a DSS
solution should be a closed curve, but for simplicity we represent it
as a fixed point. Similarly, we represent the conjectured
one-parameter family of QSS solutions (parameterised by the phase
$\delta$) as a single curve. 

Within the combined Einstein-YM-scalar system in spherical symmetry,
the pure YM critical solution then has two unstable modes and the pure
scalar field has only one. However, the phase space picture
illustrates that the QSS solution linking the two has itself only one
unstable mode, which points out of the black hole threshold towards
either collapse or dispersion. Hence we conjecture that it acts as the
critical solution for the combined Einstein-YM-scalar system,
except in the limit where the scalar field is absent.

Before doing this numerical work, we had expected that both the scalar
field and YM critical solution would be stable against perturbation by
the other field. One would then have expected there to be a third,
two-mode unstable DSS solution in which both matter fields were
present and which could be found by fine-tuning two parameters to some
$(p_*,q_*)$ with $0<q_*<1$.\footnote{Dynamically, such a solution
  $(\psi,\tilde\chi,g,G,R)=Z_{*,{\rm mixed}}(x,T)$ might have been of
  the two-oscillator form found in the context of boson stars in
  \cite{ChoptuikMasachsWay2019}, meaning that its Fourier transform in
  $T$ would show {\em sharp} peaks at multiples of two incommensurate
  scale frequencies.} The corresponding (hypothetical!)  more
conservative phase space picture is shown in
Fig.~\ref{fig:phasespace2}.

The unexpected results from our toy model highlight the question
which, if any, of those two qualitative phase space pictures is
realized for a matter field such as electromagnetism, scalar, YM or a
perfect fluid, coupled to the Einstein equations in axisymmetry.  

The only evidence that we have so far is that the spherically
symmetric critical solutions for the perfect fluid\footnote{Assuming a
  linear equation of state $P=k\rho$, for the parameter range
  $0<k\lesssim 0.49$.} \cite{Gundlachfluidnonspherical} and the scalar
field \cite{MartinGarciaGundlachscalarnonspherical} are linearly
stable to all gravitational wave perturbations.\footnote{In fact, they
  are stable against all non-spherical perturbations, except for one
  unstable $\ell=1$ perturbation associated with fluid rotation in the
  perfect fluid with $0<k<1/9$.}  Unfortunately, we do not yet have
good enough numerical simulations of axisymmetric vacuum gravitational
collapse in order to investigate any axisymmetric Einstein-matter
system in the opposite regime where the gravitational wave content
dominates the matter stress-energy, but we hope that our toy model and
phase space pictures may be helpful in the interpretation of future
results for such systems.

%%%%%%%%%%%%%%%%%%%%%%%%%%%%%%%%%%%%%%%%%%%%%%%%%%%%%%%%%%%%%%%%%%%%%%%%%%%%%%%

\acknowledgments

This research was supported through the program Research in Pairs by
the Mathematisches Forschungsinstitut Oberwolfach in 2019.  It is a
pleasure to thank the institute and its staff for hospitality during
our stay.  This work was also supported in parts by NSF grant
PHYS-1707526 to Bowdoin College, as well as through sabbatical support
from the Simons Foundation (Grant No.~561147 to TWB).  DH is supported
by the FCT (Portugal) IF Program IF/00577/2015 and PD/BD/135511/2018.
The authors would also like to acknowledge networking support by the
COST Action GWverse CA16104. 

%%%%%%%%%%%%%%%%%%%%%%%%%%%%%%%%%%%%%%%%%%%%%%%%%%%%%%%%%%%%%%%%%%%%%%%%%%%%%%%

%%%%%%%%%%%%%%%%%%%%%%%%%%%%%%%%%%%%%%%%%%%%%%%%%%%%%%%%%%%%%%%%%%%%%%%%%%%%%%%

%%%%%%%%%%%%%%%%%%%%%%%%%%%%%%%%%%%%%%%%%%%%%%%%%%%%%%%%%%%%%%%%%%%%%%%%%%%%%%%

\end{document}